\documentclass[aps,pre]{revtex4}
\usepackage{graphicx}

\begin{document}

\title{Heat capacity of square-well fluids of variable width}
\author{J. Largo}
\email{largoju@unican.es}
\author{J. R. Solana}
\email{solanajr@unican.es}
\thanks{Author to whom correspondence should be addressed}
\affiliation{Departamento de F\'{\i}sica
Aplicada, Universidad de Cantabria, 39005 Santander, Spain}
\author{L. Acedo}
\email{acedo@unex.es}
\author{A. Santos}
\email{andres@unex.es}
\affiliation{Departamento de F\'{\i}sica,
Universidad de Extremadura, 06071 Badajoz, Spain}
\date{\today}

\begin{abstract}

We have obtained by Monte Carlo NVT simulations the constant-volume excess
heat capacity of
square-well fluids for several temperatures, densities and potential
widths. Heat capacity is a
thermodynamic property much more sensitive to the accuracy of a theory than
other
thermodynamic quantities, such as the compressibility factor. This is
illustrated by
comparing the reported simulation data for the heat capacity with the
theoretical
predictions given by the Barker--Henderson perturbation theory as well as
with those
given by a non-perturbative theoretical model  based on Baxter's solution of the
Percus--Yevick integral equation for sticky hard spheres. Both theories
give accurate
predictions for the equation of state. By contrast, it is found that the
Barker--Henderson theory strongly underestimates the excess heat capacity
for low to
moderate temperatures,
  whereas a much better agreement between theory and simulation is achieved
with the
non-perturbative theoretical model, particularly for small well widths,
although the
accuracy of the latter worsens for high densities and low temperatures, as
the well width
increases.

\end{abstract}

\maketitle

\section{Introduction}

 Thermodynamic and structural properties of square-well
(SW) fluids
have been profusely
studied both from
theory and from computer simulation
%\cite{K59,BH67,SHB70,SHB71,AYM:72,T73a,T73b,T74,HBS76,HMF76,N77,SS77,HSS80,
%JKLFF81,H83,HSKGKQ84,LLA90,%
%VMRJM92,TL93,TL94a,TL94b,BYS:94,CS94,LS95,BV98,EH99,BG99,LKLLW99,NF00,DFFGSSTVZ01,
%ZFDST01,A00,VL00,AS01,RK02,LS02,VYM02,ZFDBST03}
\cite{K59}--\cite{ZFDBST03}.
{}From  the theoretical side, the first few virial coefficients have been obtained
\cite{K59,BH67,VYM02} and the radial distribution function has been
evaluated from
numerical solutions of integral equation theories, such as Percus--Yevick
\cite{T73a,T73b,T74,DFFGSSTVZ01,ZFDBST03}, Yvon--Born--Green \cite{JKLFF81}, HNC
\cite{HMF76}, MSA \cite{HSKGKQ84,DFFGSSTVZ01}, Rogers--Young
\cite{LKLLW99}, ORPA
\cite{LKLLW99}, and HRT \cite{RK02}.  Simpler analytical approximations
have also been
proposed \cite{N77,SS77,TL93,TL94a,TL94b,BYS:94,AS01}. The thermodynamic
properties have
been derived from the theoretical structure functions as well as from
perturbation
theory \cite{SHB70,SHB71,HBS76,HSS80,CS94,BG99}. Access to the ``experimental''
properties of SW fluids has been made possible via  molecular dynamics and
Monte Carlo
simulations
\cite{AYM:72,HMF76,HSS80,HSKGKQ84,LLA90,VMRJM92,EH99,LS02,ZFDBST03}.
Special
attention  has received the determination of the critical point of SW fluids
\cite{T73a,T73b,T74,HSS80,VMRJM92,CS94,BV98,EH99,NF00,VL00,AS01,RK02,ZFDBST03}, both
from the theoretical and simulational viewpoints. The main reason for this
wide interest  lies in the fact that a SW fluid is perhaps the simplest one
whose particles have attractive as well as repulsive interactions. In general,
theories are easier to apply to SW fluids than to other fluids with more
realistic
potentials. In addition, the SW potential seems to be particularly sensitive to
the performance of a theory. Therefore, this kind of fluid is an excellent
testing-ground for many theories of fluids and so the study of SW fluids can be
considered as a first step towards our understanding of the properties of
fluids with
more sophisticated interactions. There is an additional reason explaining the
recent revival of interest in SW fluids. The SW potential possesses,
besides the diameter
of the hard core and the depth of the well, an additional parameter
measuring the width
of the well. This makes the SW potential with a small width especially
suited to model
the effective interactions among colloidal particles
\cite{HSKGKQ84,LKLLW99,DFFGSSTVZ01,ZFDST01,ZFDBST03}. In this context, the glass
transition \cite{DFFGSSTVZ01,ZFDST01} and a solid-to-solid isostructural
transition
\cite{LS95} have been studied for narrow SW systems.

Despite the extensive number of studies devoted to the SW fluid, relatively
little
attention has been paid to several thermodynamic properties. This is the
case for the
heat capacity. To the best of our knowledge, there are
available \cite{AYM:72,HMF76} only a few simulation data of this property for SW
fluids. Theoretical
calculations of the same quantity are equally scarce \cite{HMF76}. In the
present paper
we have carried
out Monte Carlo simulations of the constant-volume excess heat capacity
$C_{V}^{E}$ of
SW fluids for several values of the potential width and, for each
of them, for
several densities and temperatures. Moreover, in order to put clear the
sensitivity
of  this property to the accuracy of a theory, the simulation data are
compared with the
results obtained from the
Barker--Henderson (BH) \cite{BH:67a,BH:67b} perturbation theory and with
those derived from
the theoretical model proposed by Yuste and Santos  \cite{BYS:94}, recently
simplified
by Acedo and Santos \cite{AS01}.

The paper is organized as follows. In the next section, we summarize the
theoretical
foundations of the MC procedure used and we describe the simulations
performed and the
results obtained. In Section \ref{sec3}, we present an outline of the
above-mentioned theories.
Finally, in the last section the theoretical results are compared with
simulation data and
discussed.

\section{Monte Carlo simulations}
\label{sec2}

 In a square-well (SW) fluid, particles interact by
means of a potential
of the form
\begin{equation}
u\left( r \right)=\left\{ \matrix{\infty \quad \mbox{if}\quad r\le \sigma
\hfill\cr
  -\epsilon \quad \mbox{if}\quad \sigma <r\le \lambda \sigma \hfill\cr
  0\quad \mbox{if}\quad r>\lambda \sigma \hfill\cr} \right.\label{eq:u(r)SW}
\end{equation}
 where $\lambda$ is the potential width in units of the particle
diameter $\sigma$
and $\epsilon$ is the potential depth.

 Constant-volume averaged excess heat capacity per
particle in a SW fluid
can be expressed in the form \cite{AYM:72}:
\begin{equation}
{{C_V^E} \over {Nk}}={1 \over {{{T^*}^2}}}{{\left\langle {\left( {M-\left\langle
M \right\rangle }
\right)^2} \right\rangle } \over N}, \label{eq:CVE}
\end{equation}
 where $N$ is the number of particles, $k$ is the Boltzmann constant,
$T^{*}=kT/\epsilon$ is the reduced temperature, and $M$ is the number of pairs
of interacting
particles, that is, the number of pairs of particles whose centers lie
separated by a reduced
distance $r^{*}=r/\sigma \le \lambda$.

The averages involved in equation (\ref{eq:CVE}) can be calculated by Monte
Carlo (MC)
simulations in the NVT ensemble. Therefore, we have proceeded to calculate
by means of MC
NVT the constant-volume averaged excess heat capacity per particle for SW
fluids with
well widths $\lambda$ ranging from $1.1$ to $1.5$. For each value of
$\lambda$, $C_{V}^{E}$ has
been evaluated for several densities along isotherms. To this end, a system
consisting of
$512$ particles placed in a cubic box with periodic boundary conditions was
used. Particles
were initially placed in a regular configuration and then the system was
allowed to
equilibrate for $2\times10^{4}$ cycles, each of them consisting of an
attempt move per particle, the
first $10^{4}$ cycles at a very high temperature and the remaining ones at the
desired temperature.
The calculation of $C_{V}^{E}$ was performed by averaging over the next
$5\times 10^{5}$ cycles,
performing partial averages every $10^{4}$ cycles with the aim of
estimating the statistical
error from the standard deviation.
The use of such a huge number of cycles in the
calculations was motivated by the need of ensuring that the values of
$C_{V}^{E}$ converged
to a constant value, apart from statistical fluctuations. In fact, we
realized that for
low values of the number of cycles used in the calculations, the values of
$C_{V}^{E}$
increase with the number of cycles used.

The results are shown in Table \ref{Table-CVE}. We have considered four
isotherms for
$\lambda=1.1, 1.2, 1.3$ and three isotherms for $\lambda=1.5$. The lowest
temperature in
each case is larger than the estimated critical temperature
\cite{HSS80,VMRJM92,CS94,BV98,EH99,VL00,AS01,RK02}: $T_c^*\simeq 0.5, 0.6,
0.8, 1.2$ for
$\lambda=1.1,1.2,1.3,1.5$, respectively.

\section{Theory}
\label{sec3}
\subsection{Barker--Henderson perturbation theory}

 In the second-order Barker--Henderson perturbation theory
\cite{BH:67a,BH:67b}, the free energy is expressed in the form
\begin{equation}
{F \over {NkT}}={{F_0} \over {NkT}}+{{F_1} \over {NkT}}{1 \over
{T^*}}+{{F_2} \over {NkT}}{1
\over {{{T^*}^2}}},\label{eq:FBH}
\end{equation}
 where $F_{0}$ is the free energy of the hard-sphere (HS)
reference system and
$F_{1}$ and $F_{2}$ are the first- and second-order perturbative terms,
respectively. According
to this theory, the constant-volume excess heat capacity per particle is
given by
\begin{equation}
{{C_V^E} \over {Nk}}=-{2 \over {{{T^*}^2}}}{{F_2} \over
{NkT}},\label{eq:CVE-PER}
\end{equation}
 where
\begin{equation}
{{F_2} \over {NkT}}=-\pi \rho kT\left( {{{\partial \rho } \over {\partial P}}}
\right)_0\int_0^\infty  {\left[ {u_1^*\left( r \right)} \right]^2g_0\left( r
\right)r^2dr}
\label{eq:F2mc}
\end{equation}
 in the so-called {\em macroscopic compressibility
approximation}, whereas
\begin{equation}
{{F_2} \over {NkT}}=-\pi \rho kT\int_0^\infty  {\left[ {u_1^*\left( r \right)}
\right]^2\left\{ {{{\partial \left[ {\rho g_0\left( r \right)} \right]}
\over {\partial P}}} \right\}_0r^2dr}\label{eq:F2lc}
\end{equation}
 in the so-called {\em local compressibility approximation}. In Eqs.\
(\ref{eq:F2mc}) and (\ref{eq:F2lc}), $\rho=N/V$ is the number density,
$u_{1}^{*}(r)=u_{1}(r)/\epsilon$ is the
perturbative contribution to the potential function, which in a SW potential is
$u_{1}^{*}(r)=-1$ for $\sigma<r<\lambda\sigma$,
$P$ is the pressure, and $g_{0}(r)$ is the radial distribution function
(r.d.f.) of the
hard-sphere reference fluid.

In recent years, several analytical and very accurate expressions for the
r.d.f. $g_{0}(r)$ of
the HS fluid have been proposed \cite{BS:91,BLS:96,TL:95}. They can be used
to determine
$F_{2}$ in expressions (\ref{eq:F2mc}) and (\ref{eq:F2lc}). Regarding
$(\partial \rho / {\partial P})_{0}$, which appears explicitly in expression
(\ref{eq:F2mc}) and implicitly in (\ref{eq:F2lc}),  it can be obtained from the
well-known
Carnahan--Starling \cite{CS:69} equation of state
\begin{equation}
Z_0={{P_0V} \over {NkT}}={{1+\eta +\eta ^2-\eta ^3} \over {\left( {1-\eta }
\right)^3}},
\label{eq:Z-CS}
\end{equation}
 where $\eta=(\pi/6)\rho\sigma^{3}$ is the packing fraction.

\subsection{Yuste--Acedo--Santos model}

 The internal energy can be obtained from the r.d.f
$g(r)$ through
the energy equation
\begin{equation}
U={3 \over 2}NkT+2\pi N\rho \int_0^\infty  {u\left( r \right)g\left( r
\right)r^2dr},
\label{eq:EE}
\end{equation}
 whence
\begin{equation}
{{C_V^E} \over {Nk}}={{2\pi \rho } \over k}\int_0^\infty  {u\left( r
\right)\left[
{{{\partial g\left( r \right)} \over {\partial T}}} \right]_Vr^2dr}.
\label{eq:CVE-EE}
\end{equation}
In the special case of the SW potential (\ref{eq:u(r)SW}), Eqs.\
(\ref{eq:EE}) and (\ref{eq:CVE-EE}) become
\begin{equation}
U={3 \over 2}NkT-12N\epsilon\eta \int_1^\lambda  {g\left( r^*
\right){r^*}^2dr^*},
\label{eq:EE_SW}
\end{equation}
\begin{equation}
{{C_V^E} \over {Nk}}=-12\eta \int_1^\lambda  \left[
{\partial g\left( r^* \right)} \over {\partial T^*} \right]_\eta {r^*}^2dr^*,
\label{eq:CVE-EE_SW}
\end{equation}
respectively.
The r.d.f $g(r^*)$ of the SW
fluid depends on the packing fraction $\eta$, the reduced temperature $T^*$
and, parametrically, on the well width $\lambda$. In principle, one has to
resort to numerical solutions of integral equation theories. On the other
hand,
particularly
suitable for the
purpose of obtaining the heat capacity is the  heuristic model proposed by
Yuste and Santos \cite{BYS:94} and recently simplified by Acedo and Santos
\cite{AS01},
which is analytical
and fairly accurate. Henceforth we will refer to this model as the
Yuste--Acedo--Santos (YAS) model. It is based on expressing the Laplace
transform $G(t)$ of $r^*g(r^*)$ in the
form
\begin{equation}
G\left( t \right)=t{{F\left( t \right)e^{-t}} \over {1+12\eta F\left( t
\right)e^{-t}}}=\sum\limits_{n=1}^\infty  {\left( {-12\eta }
\right)^{n-1}t\left[
{F\left( t \right)} \right]^ne^{-nt}},
\label{eq:G(t)}
\end{equation}
 where the auxiliary function $F(t)$ is assumed to have the form
\cite{BYS:94,AS01}
\begin{equation}
F(t)=-\frac{1}{12\eta}\frac{e^{1/T^*}+K_1 t-\left(e^{1/T^*}-1+K_2
t\right)e^{-(\lambda-1)t}}{1+S_1 t+S_2 t^2+S_3 t^3}.
\label{eq:F(t)}
\end{equation}
The coefficients $K_1$, $K_2$, $S_1$, $S_2$ and $S_3$ are {\em explicit}
functions of $\eta$, $T^*$ and $\lambda$
determined from consistency conditions. We refer the interested reader
to Refs.\
\cite{BYS:94,AS01} for further details.
The YAS model (\ref{eq:F(t)}) reduces to the exact solutions of the
Percus--Yevick (PY) equation in the limit of hard spheres ($\lambda\to 1$
or $T^*\to\infty$) \cite{W63,T63}, as well as in the limit of sticky hard
spheres ($\lambda\to 1$ and $T^*\to 0$ with $T^*\sim -1/\ln(\lambda-1)$)
\cite{B68}. {}From that point of view, the approximation (\ref{eq:F(t)})
can be seen as a simple extension to finite widths of Baxter's solution of
the PY equation for sticky hard spheres.

Upon Laplace inversion of Eq.\
(\ref{eq:G(t)}),
the final expression of the r.d.f. reads
\begin{equation}
g\left( {r^*} \right)={r^*}^{-1}\sum\limits_{n=1}^\infty  {\left( {-12\eta }
\right)^{n-1}f_n\left( {r^*-n} \right)\Theta \left( {r^*-n} \right)},
\label{eq:g(r)-SW}
\end{equation}
 where the functions $f_{n}(r^{*})$ are the inverse Laplace
transforms of
$t[F(t)]^{n}$ and
$\Theta (r^*)$ is  Heaviside's step function.
Therefore, to
determine the r.d.f. for $r^* < n+1$ only the first $n$ terms in the summation
(\ref{eq:g(r)-SW}) are needed.
In particular, for the values of $\lambda\leq 2$ considered in
this paper, one has
\begin{equation}
g(r^*)=-\frac{{r^*}^{-1}}{12\eta}\sum_{i=1}^3z_i\frac{e^{1/T^*}+K_1
z_i}{S_1+2S_2 z_i+3S_3 z_i^2}e^{z_i(r^*-1)}, \quad 1<r^*\leq \lambda,
\label{1}
\end{equation}
where $z_i$ ($i=1,2,3$) are the three roots of the cubic equation $1+S_1
t+S_2 t^2+S_3 t^3=0$. Inserting Eq.\ (\ref{1}) into Eq.\
(\ref{eq:CVE-EE_SW}), we finally get
\begin{equation}
{{C_V^E} \over {Nk}}=\frac{\partial}{\partial
T^*}\sum_{i=1}^3\frac{e^{1/T^*}+K_1 z_i}{S_1+2S_2 z_i+3S_3
z_i^2}\left[z_i^{-1}-1+(\lambda-z_i^{-1})e^{z_i(\lambda-1)}\right].
\label{2}
\end{equation}

The heat capacity can also be obtained from the YAS r.d.f. by following the
virial and compressibility routes to the equation of state. The reason for
the choice of the energy route (\ref{eq:EE}) is two-fold. First, it is
obviously the most direct route to determine the heat capacity. Second, we
have checked that the other routes yield results that present larger
deviations from the simulation data. This latter observation is consistent
with the case of the PY theory for sticky hard spheres \cite{WHB71} and for
SW fluids
\cite{T73b,T74}.

\section{Results and discussion}

\begin{figure}[h!]
\begin{center}
\includegraphics[width=15cm]{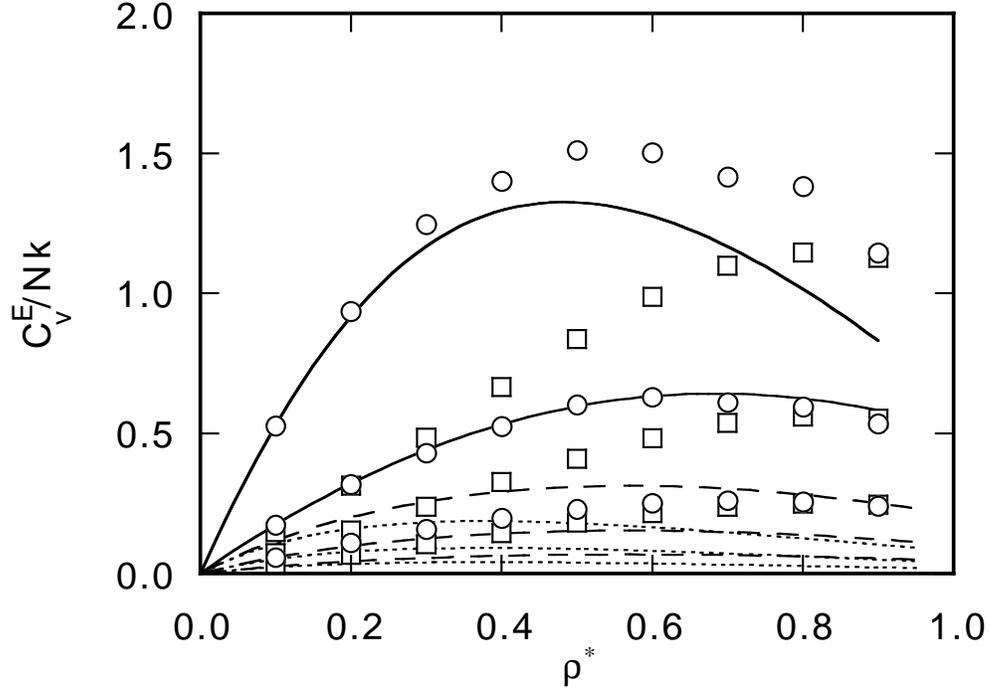}
\vspace{-5 cm}
\caption{Constant-volume excess heat capacity for a SW fluid with
$\lambda=1.1$
as a function of the reduced density $\rho^{*}$. Circles:
simulation data from Table \ref{Table-CVE} for $T^{*}=0.7$, $T^{*}=1.0$,
and $T^{*}=1.5$,
respectively, from top to down. Squares: values obtained from Eq.
(\ref{eq:CVE-PER})
using the simulation data of $F_{2}$ reported in \cite{LS:03}. Continuous
curve: YAS model.
Dashed curve: BH perturbation theory in the
local compressibility approximation. Dotted curve:  BH perturbation theory
in the
macroscopic compressibility approximation.}
\label{Cv11}
\end{center}
\end{figure}

\begin{figure}[h!]
\begin{center}
\includegraphics[width=15 cm]{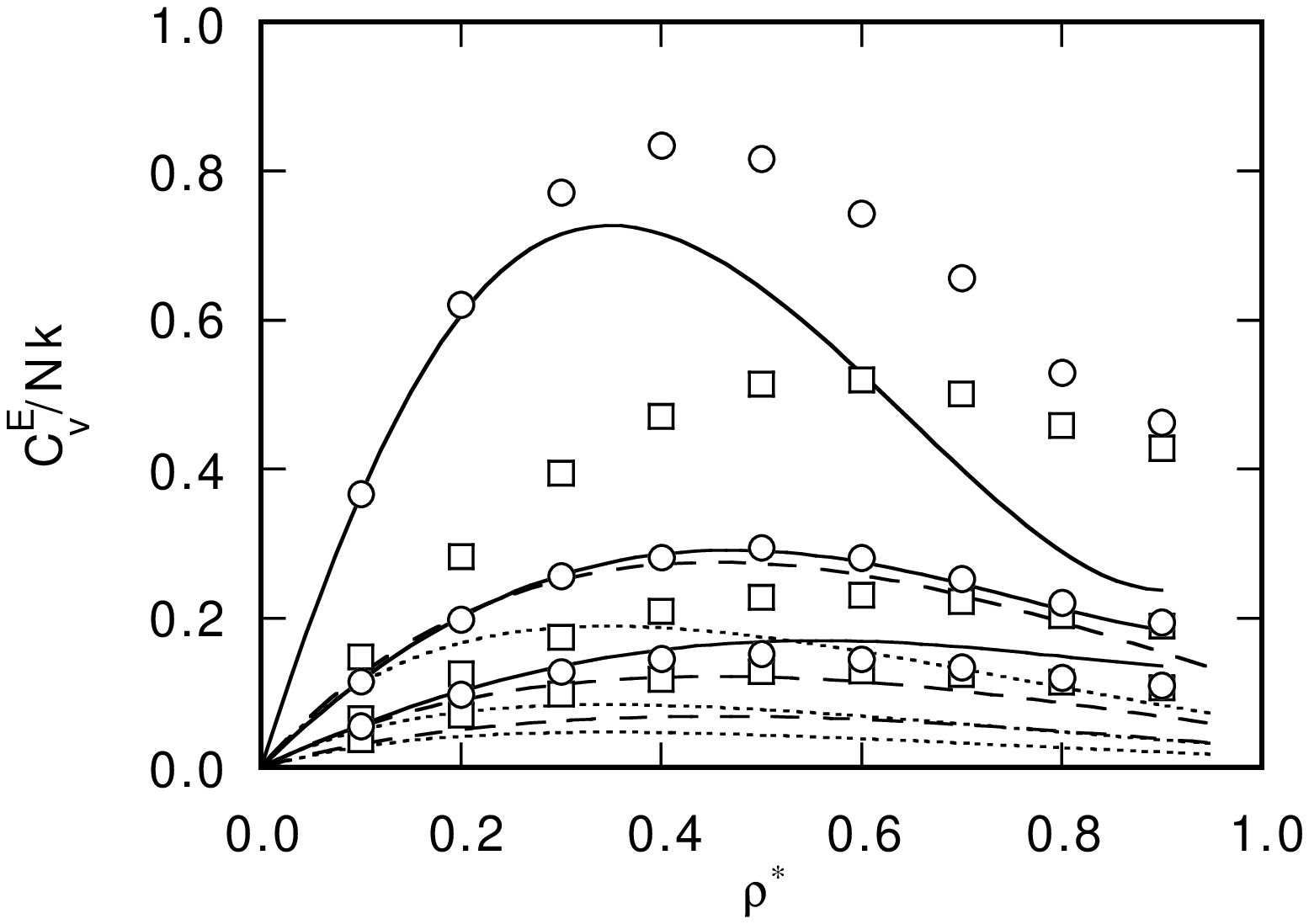}
\vspace{-4 cm}
\caption{As in Fig.\ \ref{Cv11} for $\lambda=1.2$, except that the
temperatures are
$T^{*}=1.0$, $T^{*}=1.5$, and $T^{*}=2.0$, respectively, from top to down.}
\label{Cv12}
\end{center}
\end{figure}

\begin{figure}[h!]
\begin{center}
\includegraphics[width=15 cm]{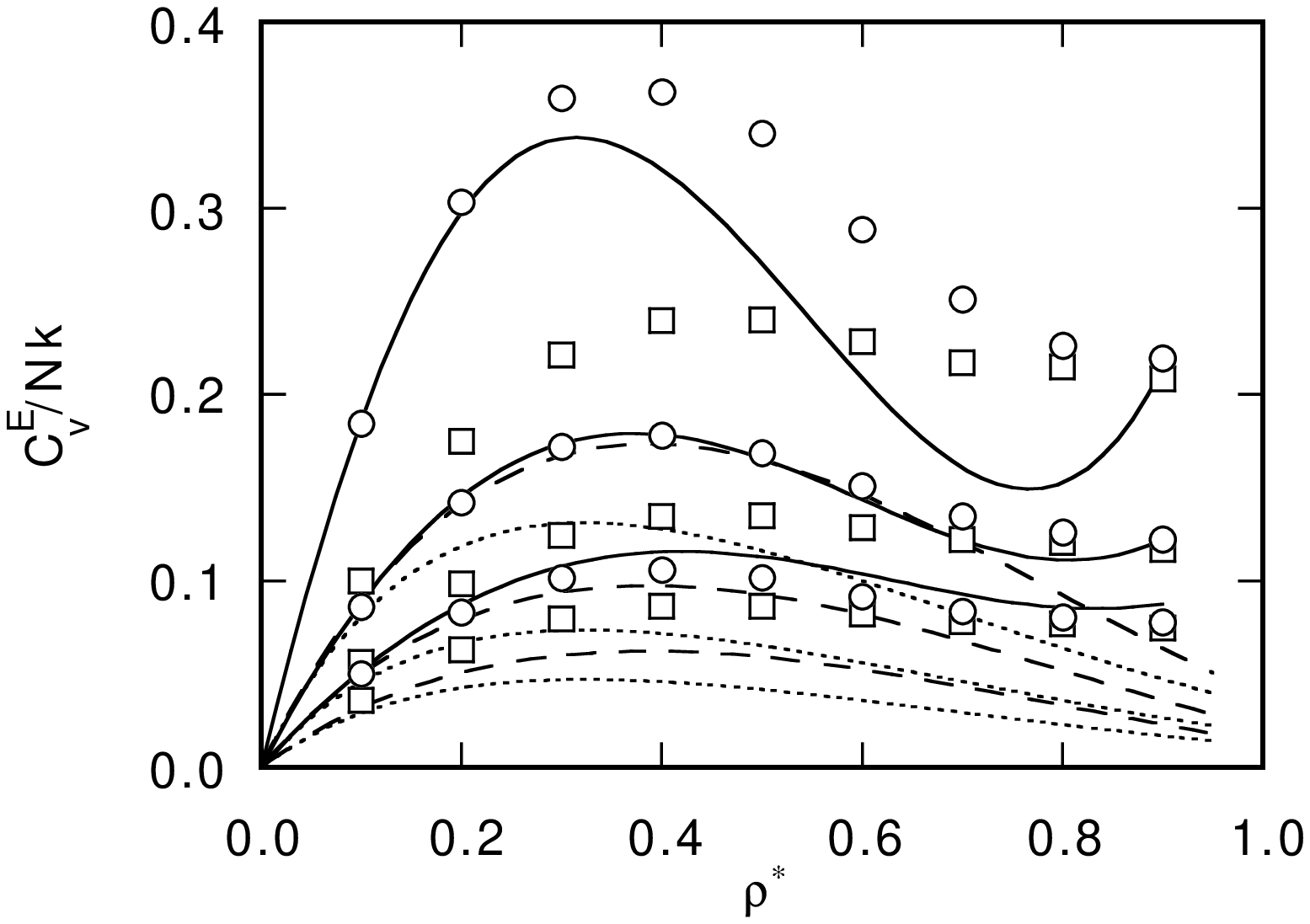}
\vspace{-4 cm}
\caption{As in Fig. \ref{Cv11} for $\lambda=1.3$, except that the
temperatures are
$T^{*}=1.50$, $T^{*}=2.0$, and $T^{*}=2.5$, respectively, from top to down.}
\label{Cv13}
\end{center}
\end{figure}

\begin{figure}[h!]
\begin{center}
\includegraphics[width=15 cm]{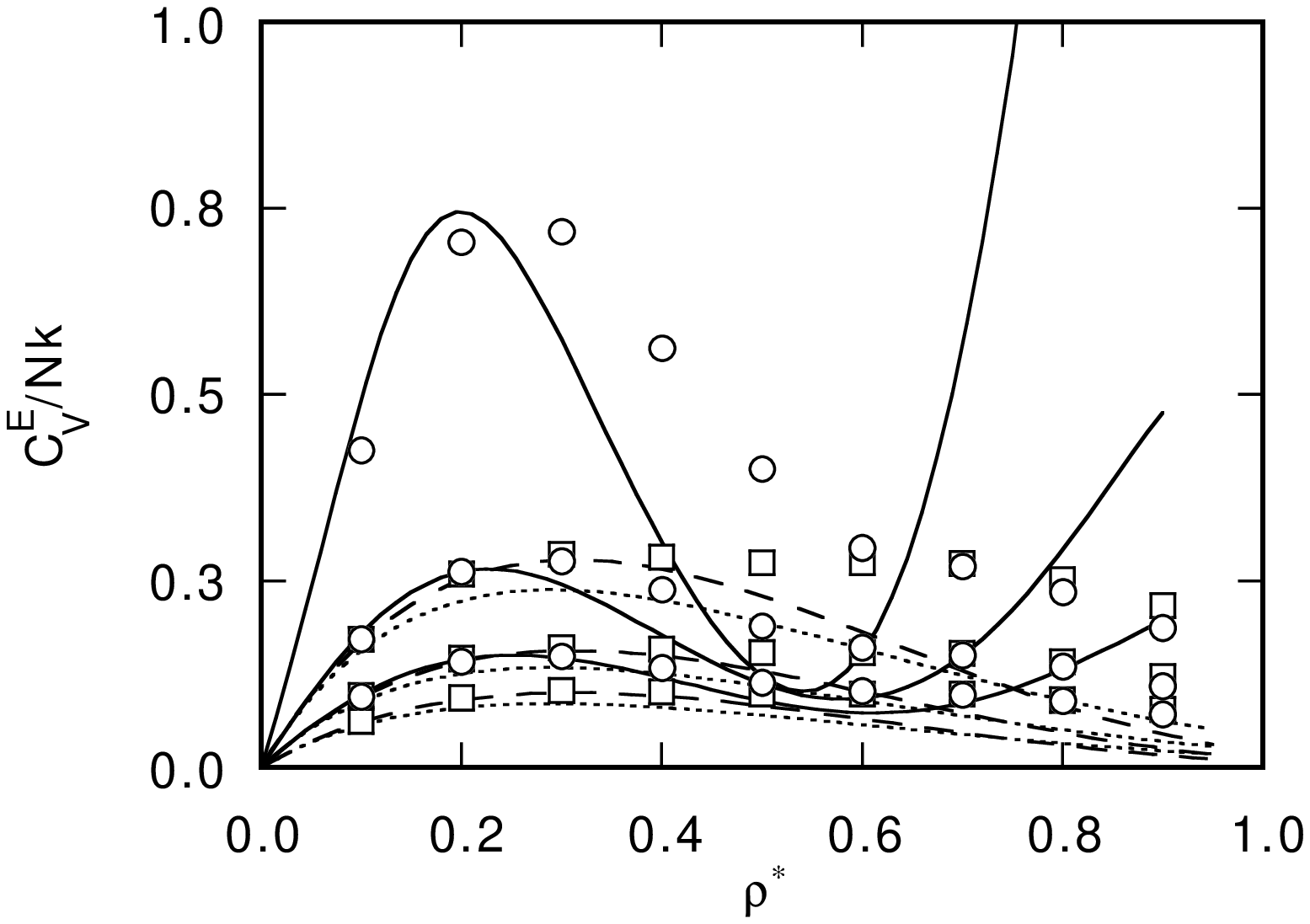}
\vspace{-4 cm}
\caption{As in Fig. \ref{Cv11} for $\lambda=1.5$, except that the
temperatures are
$T^{*}=1.50$, $T^{*}=2.0$, and $T^{*}=2.5$, respectively, from top to down.}
\label{Cv15}
\end{center}
\end{figure}

Results obtained for $C_{V}^{E}$ from the
second-order Barker--Henderson perturbation
theory within the
local compressibility approximation as well as within the macroscopic
compressibility
approximation are compared in Figs.\ \ref{Cv11}--\ref{Cv15} with the simulation
data of Table
\ref{Table-CVE}. We can see that although the local compressibility
approximation provides a
better agreement with simulation data, both approximations are rather poor
at low
temperatures. This might be due either to the insufficient accuracy of both
the local
compressibility and the macroscopic compressibility approximations or to
the fact that
higher order terms, beyond the second one, in the expansion of the
Helmholtz free energy
in power series of the  inverse of the reduced temperature, have a nonnegligible
contribution to the heat capacity. In order to determine which of these two
possibilities
is the right one, we can use for $F_{2}$ in Eq. (\ref{eq:CVE-PER})
simulation data, thus
avoiding theoretical approximations. These simulations were performed by
Barker and
Henderson \cite{BH:72} who reported the results in terms of a function
depending on forty
five parameters for each density. These parameters were determined from a least
squares fitting of their simulation data. Since the use of that fitting is
somewhat
tedious, we have preferred to use directly simulation data for $F_2$, which
are available
for several densities and well widths \cite{LS:03}, to determine
$C_{V}^{E}$ from Eq.
(\ref{eq:CVE-PER}). As one can see in Figs.\ \ref{Cv11}--\ref{Cv15},
results thus
obtained are much closer to the theoretical results derived
from the BH second-order perturbation theory than to the values of
$C_{V}^{E}$ obtained
from direct simulations, except in the high density limit. This means that
the main reason
of the failure of the Barker-Henderson perturbation theory in predicting the
heat capacity of SW fluids arises in the truncation of the perturbative
series at the level
of the second order term, the higher order terms having a nonnegligible
contribution. This
is in contrast with the situation for the equation of state
\cite{BH:67a,LS:XX}, which
is accurately given by the second-order BH perturbation theory even at
relatively low
temperatures for
wide ranges of densities and potential wells. The reason is that, as
pointed before,
the constant-volume excess heat capacity is a thermodynamic property
particularly sensitive
to the performance of a theory and therefore the influence of higher order
terms, which is
small in the equation of state, may be important in the heat capacity. This
is particularly
true for low values of the potential width, since the lower the potential
width, the
slower the convergence of the BH perturbation theory  at
low temperatures \cite{H83}.

A much better agreement is obtained with YAS theory, Eq.\  (\ref{2}), at
low to moderate
densities, as shown in the same figures. This theory, in contrast  to the
BH theory,
provides a  better agreement with the simulation data of $C_{V}^{E}$ as
the potential
width decreases. This is consistent with the fact that,  as said before,
the YAS model is
an extension to $\lambda>1$ of the PY solution for sticky hard spheres and
hence it is
expected to be as accurate as the PY theory at least for small $\lambda-1$.  The
structural properties predicted by the YAS model  for the SW fluid also
exhibits a good
agreement with simulation data for low values of $\lambda-1$ whereas the
accuracy worsens
as $\lambda$ increases \cite{BYS:94,AS01}. Figures \ref{Cv11}--\ref{Cv15}
show that, given
a well width $\lambda$, the YAS values of $C_{V}^{E}$ are more accurate as the
temperature increases and/or the density decreases.

In conclusion, we have performed Monte Carlo simulations of the
constant-volume excess
heat capacity of SW fluids of variable width for a wide range of densities
and at
several characteristic temperatures. This thermodynamic quantity vanishes
for hard
spheres and so it represents an important measure of the influence of
attractive forces
on the state of the fluid. Moreover, the heat capacity seems to provide a rather
stringent test to assess the accuracy of theoretical approaches. In this
paper we have
compared the simulation data with the Barker--Henderson perturbation theory
\cite{BH:67a,BH:67b} and with a non-perturbative theory developed by Yuste,
Acedo, and
Santos
\cite{BYS:94,AS01}. While the former theory presents a poor performance,
which can be
attributed to the truncation of the perturvative series to second order
rather than the
inaccuracy of the theory itself, the  non-perturbative theory does a fairly
good job,
especially for narrow wells, except  at low temperatures and high
densities. Although a
potential well of $\lambda=1.5$ is appropriate for many simple fluids, SW
fluids with
lower values of $\lambda$ may be of interest because the properties of certain colloidal
suspensions are well reproduced by considering SW interactions with narrow
potential
widths. Therefore, as several theories for SW fluids have achieved a
considerable accuracy
for the equation of state and the pair correlation function of SW fluids,
the constant
volume excess heat capacity seems to be a suitable thermodynamic property
to discriminate
between them. In this context, we expect that our simulation data can
stimulate other
studies on the heat capacity of SW fluids of variable width and can be used
to check the
reliability of other approximations.

\vspace{1 cm}

The present work has been partially supported by the Spanish Direcci\'{o}n
General de
Investigaci\'{o}n (DGI) under grants No.\ BFM2000-0014 (J.L. and J.R.S) and
No.\
BFM2001-0718 (L.A and A.S.).

\pagebreak

%\section*{List of tables}

\begin{table}[h!]
\vspace{-3 cm}
\caption{MC simulation data for $C_{V}^{E}/Nk$. The numbers enclosed between
parentheses indicate the
statistical uncertainty in the last decimal places.}
\begin{center}
\begin{tabular}{lccccc}
\hline
   $\rho^{*}$   &    $T^*=0.7$    &    $T^*=1.0$ &    $T^*=1.5$ &
$T^*=2.0$   &
$T^*=2.5$  \\

\hline

\multicolumn{6}{c}{$\lambda=1.1$}\\

 0.1           &   0.527(3) &     0.1731(8) &    0.0580(2)   &        &     \\
 0.2           &   0.94(1) &     0.319(1) &    0.1109(3)   &    0.0556(2)
&     \\
 0.3           &   1.25(2)&     0.431(3) &    0.1588(8)   &       &     \\
 0.4           &   1.40(2) &     0.525(5) &    0.1979(9)   &    0.1033(4)
&     \\
 0.5           &   1.51(3) &     0.603(7) &    0.230(1)   &       &     \\
 0.6           &   1.50(3) &     0.630(7) &    0.252(2)   &    0.1361(7)
&     \\
 0.7           &   1.42(3) &     0.612(7) &    0.261(3)   &       &     \\
 0.8           &   1.38(2) &     0.595(6) &    0.256(3)   &    0.142(1)
&     \\
 0.9           &   1.15(3) &     0.534(7) &    0.241(4)   &       &     \\
\hline
\multicolumn{6}{c}{$\lambda=1.2$}\\

 0.1           &   							     &     0.367(2) &    0.1155(3)   &
0.0559(1)       &     \\
 0.2           &   3.11(17) &     0.621(7) &    0.1988(6)   &    0.0979(3)
&     \\
 0.3           &   							     &     0.77(1) &    0.2568(9)   &
0.1288(6)       &     \\
 0.4           &   3.68(22) &     0.83(1) &    0.282(2)   &    0.1463(6)
&     \\
 0.5           &                                       &     0.82(1) &
0.295(2)   &    0.1528(6)       &     \\
 0.6           &   3.01(14) &     0.743(9) &    0.282(3)   &    0.1459(8)
&     \\
 0.7           &                                       &     0.657(8) &
0.253(2)   &    0.1351(7)       &     \\
 0.8           &   1.35(5) &     0.530(8) &    0.222(2)   &    0.1209(9) \\
 0.9           &                                       &     0.463(6) &
0.195(2)   &    0.111(1)       &     \\
\hline
\multicolumn{6}{c}{$\lambda=1.3$}\\

 0.1              &      &                                     &
0.1846(6)   &    0.0864(2)   &  0.0504(1) \\
 0.2              &      &  1.17(1) &    0.303(2)   &    0.1423(7)   &
0.0834(3) \\
 0.3              &      &                                     &
0.359(4)   &    0.1720(5)   &  0.1016(4) \\
 0.4              &      &  1.35(2) &    0.363(2)   &    0.178(1)   &
0.1060(6) \\
 0.5              &      &                                     &
0.340(3)   &    0.1688(8)   &  0.1019(7) \\
 0.6              &      &  0.90(2) &    0.289(3)   &    0.151(1)   &
0.0918(3)\\
 0.7              &      &                                     &
0.251(2)   &    0.1350(9)   &  0.0838(5) \\
 0.8              &      &  0.512(8) &    0.226(2)   &    0.126(1)   &
0.0807(6)\\
 0.9              &      &                                     &
0.219(2)   &    0.122(1)   &  0.078(1) \\
\hline
\multicolumn{6}{c}{$\lambda=1.5$}\\

 0.1           &      &                               &    0.426(3)   &
0.1724(4)  &    0.0952(3)  \\
 0.2           &      &             &    0.705(5)   &    0.263(2)  &
0.1426(8)  \\
 0.3           &      &                               &    0.719(9)   &
0.277(2)  &    0.1495(7)  \\
 0.4           &      &             &    0.563(5)   &    0.239(2)  &
0.1339(8) \\
 0.5           &      &                               &    0.401(6)   &
0.190(2)  &    0.1142(6)  \\
 0.6           &      &             &    0.295(2)   &    0.161(1)  &
0.1027(5) \\
 0.7           &      &                               &    0.270(2)   &
0.1513(6)  &    0.0977(6)  \\
 0.8           &      &             &    0.235(3)   &    0.136(1)  &
0.0894(9) \\
 0.9           &      &                               &    0.188(2)   &
0.109(2)  &    0.0716(6)  \\

\hline
\end{tabular}
\end{center}
\label{Table-CVE}
\end{table}

\clearpage

\section*{List of figures}

%\pagebreak

\end{document}